\documentstyle[preprint,aps,epsf,floats]{revtex}

\begin{document}

\draft


\title{Atom Chips}
\author{Ron Folman$^1$, Peter Kr\"uger$^1$, Donatella Cassettari$^1$, Bj\"orn
Hessmo$^{1,2}$, Thomas Maier$^3$, J\"org Schmiedmayer$^1$}

\address{$^1$Institute f\"ur Experimentalphysik,
Universit\"at Innsbruck, A-6020 Innsbruck,  Austria\\
$^2$Department of Quantum Chemistry, Uppsala University, S-75120
Uppsala, Sweden\\
$^3$Institute f\"ur Festk\"orperelektronik, Floragasse 7, A-1040 Wien, Austria}

\date{\today}

\maketitle

\begin{abstract}
Atoms can be trapped and guided using nano-fabricated wires on
surfaces, achieving the scales required by quantum information
proposals. These Atom Chips form the basis for robust and
widespread applications of cold atoms ranging from atom optics to
fundamental questions in mesoscopic physics, and possibly quantum
information systems.
\end{abstract}


\vskip1pc


In mesoscopic quantum electronics, electrons move {\em inside}
semiconductor structures and are manipulated using potentials where
at least one dimension is comparable to the de-Broglie wavelength
of the electrons \cite{imry,moti}. Similar potentials can be
created for neutral atoms moving microns {\em above} surfaces,
using nano-fabricated charged and current carrying structures
\cite{schm,joerg2,hind}. Surfaces carrying such structures form
Atom Chips which, for coherent matter wave optics, may form the
basis for a variety of novel applications and research tools,
similar to what integrated circuits are for electronics.

In this work we make use of the magnetic interaction $V_{mag} =
-\vec{\mu} \cdot \vec{B}$ based on the coupling of the atomic
magnetic moment $\vec{\mu}$ to the magnetic field $\vec{B}$ to
trap and manipulate atoms close to the surface of an Atom Chip.
The trapping potentials are created by superposing a homogeneous
magnetic bias field with the field generated by a thin current
carrying wires. The trap depth is given by the homogeneous field,
the gradients and curvatures by the magnetic fields from the wire
\cite{joerg2,joerg1}.

We have previously reported on the manipulation of neutral atoms
using thin (down to below $1 \mu m$) charged wires \cite{Den98}
and current carrying wires (down to $25 \mu m$) to form guides
\cite{joerg1,Schm92}, beam splitters \cite{joerg2}, and Z or U
shaped 3-dimensional traps \cite{joerg3}. These structures were
free standing.

The next step was to turn to surface mounted wires
\cite{SurfaceMounting} which was recently achieved for large
structures \cite{fortagh,surface}. However,the full potential in
surface mounted atom optics lies in the robust miniaturization
down to the mesoscopic scale. Such a move is primarily motivated
by the theoretically required scale needed to achieving
entanglement with neutral atoms through controlled collisions
\cite{zoller} or cavity QED \cite{QED}, entanglement being the
basic building block for quantum information devices.

Here we present such a nanofabricated device with which the
required ground state size of less than $100nm$ was achieved. This
is a first step towards our vision, the realization of a fully
integrated {\em Atom Chip}. We start by describing the chip and
the experimental setup, followed by a presentation of the results.
Finally, we discuss potential applications and future
perspectives.

The chip we have used in this work is made of a $2.5 \, \mu m$
gold layer placed on a $600\mu m$ thick GaAs substrate
\cite{maier}.  The gold layer is patterned using nano-fabrication
technology.  The scale limit of the process used is well below
$100 \, nm$.

In figure 1a we present the main elements of the chip design used
in the work described here. Each of the large U-shaped wires,
together with a bias field, creates a quadrupole field, which may
be used to form a Magneto-Optical-Trap (MOT) on the chip as well
as a magnetic trap. Both U-shaped wires together may be used to
form a strong magnetic trap in order to 'load' atoms into the
smaller structures, or as an on-board (i.e. without need for
external coils) bias field, for guides and traps created by the
thin wire running between them. The thin wires are $10\mu m$ wide
and depending on the contact used, may form a U-shaped or a
Z-shaped magnetic trap or a magnetic guide. The chip wires are all
defined by boundaries of $10 \, \mu m$ wide etchings in which the
conductive gold has been removed.  This leaves the chip as a gold
mirror (with $10 \, \mu m$ etchings) and it can be used to reflect
the laser beams for the MOT during the cooling and collecting of
atoms. Figure 1b presents the mounted chip before it is introduced
into the vacuum chamber.  In addition, a U-shaped $1 \, mm$ thick
wire, capable of carrying up to $20 \, A$ of current, has been put
underneath the chip in order to assist with the loading of the
chip. Its location and shape are identical to those of one of the
$200\mu m$ U-shaped wires and it differs only in the amount of
current it can carry.

The chip assembly (Fig. 1b) is then mounted inside a vacuum chamber
used for atom trapping experiments, with optical access for the
laser beams and the observation cameras and with the possibility of
applying the desired bias fields (Fig. 2).  For a more detailed
description of the apparatus and the atom trapping procedure, see
\cite{joerg2,joerg1,denschlag}.

The experimental procedure for loading cold atoms into the small traps on the
chip is the following:

In the first step typically $10^8$ $^{7}Li$ atoms are loaded from
an effusive atomic beam into a MOT \cite{LiMOT}. Because the atoms
have to be collected a few millimeters away from a surface we use
a 'reflection' MOT \cite{reflect}. Thereby, the 6 laser beams
needed for the MOT are formed from 4 beams by reflecting two of
them off the chip surface (Fig.~\ref{fig:pict2}). Hence atoms
above the chip actually encounter six light beams. To assure a
correct magnetic field configuration needed for the formation of a
MOT, one of the reflected light beams has to be in the axis of the
MOT coils. Figure 3a shows a top view of the chip and the
reflection MOT sitting above the U-shaped wires.

The large external quadrupole coils are then switched off while
the current in the U-shaped wire underneath the chip is switched
on (up to $16A$), together with an external bias field ($8G$).
This forms a nearly identical, but spatially smaller, quadrupole
field as compared to the fields of the large coils.  The atoms are
thus transferred to a secondary MOT which by construction is
always well aligned with the chip (Fig. 3b).  By changing the bias
field, the MOT can be shifted close to the chip surface
(typically, $2 \, mm$). The laser power and detuning are changed
to further cool the atoms, giving us a sample with a temperature
below $200 \, \mu K$.

In the next step, the laser beams are switched off and the
quadrupole field serves as a magnetic trap in which the low field
seeking atoms are attracted to the minimum of the field.  Without
the difficulties of near surface shadows hindering the MOT, the
magnetic trap can now be lowered further towards the surface of
the chip (Fig. 3c). This is simply done by increasing the bias
field (up to $19 \, G$).  Atoms are now close enough so that they
can be trapped by the chip fields. The loading of the chip has
begun.

Next, 2A are sent through each of the two $200 \, \mu m$ U-shaped
wires on the chip and the current in the U-shaped wire located
underneath the chip is ramped down to zero.  This procedure brings
the atoms even closer to the chip, compresses the trap
considerably, and transfers the atoms to a magnetic trap formed by
the currents in the chip. The distances of the atoms from the
surface are now typically a few hundred microns (Fig. 3d).

Finally, the $10 \, \mu m$ wire trap is loaded in much the same
way. It first receives a current of $300 \, mA$.  Then the current
in both the U-shaped wires is ramped down to zero
(Fig.~\ref{fig:compression}). Atoms are now typically a few tens
of microns above the surface (Fig. 3e).


These guides and traps can be further compressed by ramping up the
bias magnetic field.  In this process we typically achieve
gradients of $>25 \, kG/cm$.  By applying a bias field of $40 \,
G$ and a current of $200 \, mA$ in the $10 \, \mu m$ wire we
achieve trap parameters with a transverse ground state size below
100 nm and frequencies of above $100 \, kHz$ (as required by the
quantum computation proposals \cite{zoller}).

By running the current through a longer $10\mu m$ section of the
thin wire, we turn the magnetic trap into a guide, and atoms could
be observed expanding along it (Fig. 3f).

In an additional experiment we used the thick wires on the chip to
create an {\em on chip} bias fields for the trapping. In the
experiment this is done by sending current through the two U-traps
in the opposite direction with respect to the current in the $10
\, \mu m$ wire, which creates a magnetic field parallel to the
chip surface. Hence, we demonstrate trapping of atoms on a self
contained chip.

In these small traps, the atom gas can be compressed to the point
where direct visual observation is difficult.  In such a case, we
observe those atoms after guiding or trapping,  by 'pulling' them up
from the surface into a less compressed wire trap (by increasing the
wire current or decreasing the bias field).

During the transfer from the large magnetic trap to the small 10
$\mu m$ trap the density of the atomic cloud is increased by up to
a factor 350. As the trap is compressed, the temperature of the
atoms rises, and if in this course the trapping potential is not
deep enough atoms are lost. In our case, the trap depth is
uniquely determined by the bias field used, which leads to depths
$E=-m_Fg_F\mu_B|B|$ ranging  between $ \sim 6 \, MHz$ ($ \sim 0.25
\, mK$) for the $8G$ bias field and $|m_F|=1$ to $\sim 70 \, MHz$
($ \sim 3 \, mK$) for the $50G$ bias field and $|m_F|=2$.  This
adiabatic heating and the finite trap depth limited the transfer
efficiency for atoms from the large magnetic quadrupole into the
smallest chip trap to $<$50 \%.

Since we use an trapped atomic sample consisting of 3 different
spin states ($|F=2, m_F=2\rangle$, $|F=2, m_F=1\rangle$, and
$|F=1, m_F=-1\rangle$) the large compression also increases the
rate for inelastic two body spin flip collisions dramatically.
This rate is for our Li sample similar to the elastic collision
rate \cite{cote} and is therefore a good estimate of the
achievable collision rates in a polarized sample. From measured
decay curves we estimate the collision rate to be in the order of
20 $s^{-1}$ for atoms in a typical small chip trap.  This estimate
of the scattering rate in the small chip traps is supported by the
observation that the atoms expand very fast into the wire guide,
indicating that energy gained from the transverse compression of
the trap is transformed efficiently into longitudinal velocity at
a very high rate.

The above shows that the concept of an Atom Chip clearly works. We
have demonstrated that a wide variety of magnetic potentials may
be realized with simple wires on surfaces. Wires together with a
bias field can produce quadrupole fields for a MOT, 3D minima for
trapping, and 2D minima for guiding. Furthermore it is very easy
to manipulate the center of the trap and its width.  We have shown
that loading such an atom trap $\mu m$ above the surface does not
present a major problem and trap parameters with a transverse
ground state size below $100 \, nm$ and frequencies of above $100
\, kHz$ have been achieved. In addition we could trap atoms
exclusively with the chip fields, creating the required bias
fields 'on board'. Last but not least, it has been shown that
standard nano-fabrication techniques and materials may be utilized
to build these Atom Chips.  The wires on the surface can stand
sufficiently high current densities ($>10^6 \, A/cm^2$) in vacuum
and at room temperature.  Together with the scaling laws of these
traps \cite{joerg2,joerg1,joerg3}, this will allow us to use much
thinner wires and reach traps with ground state sizes of $10 \,
nm$ and trap frequencies in the MHz range.

We conclude with a long term outlook.  In this work we have
successfully realized a step which is but one of many still needed.
A final integrated Atom Chip, should have a reliable source of cold
atoms, for example a BEC \cite{BEC}, with an efficient loading
mechanism, single mode guides for coherent transportation of atoms,
nano-scale traps, movable potentials allowing controlled collisions
for the creation of entanglement between atoms, extremely high
resolution light fields for the manipulation of individual atoms,
and internal state sensitive detection to read out the result of the
processes that have occurred (e.g. the quantum computation). All of
these, including the bias fields and probably even the light
sources, could be on-board a self-contained chip.  This would
involve sophisticated 3D nano-fabrication and the integration of a
diversity of electronic and optical elements, as well as extensive
research into fundamental issues such as decoherence near a surface.
Such a robust and easy to use device, would make possible advances
in many different fields of quantum optics: from applications in
atom optics \cite{AtomOptics} such as clocks and sensors to
implementations of quantum information processing and communication
\cite{QIPC}.

We would like to thank  A. Chenet, A. Kasper and A. Mitterer for
help in the experiments.  Atom chips used in the preparation of
this work and in the actual experiments were fabricated at the
Institut f\"{u}r Festk\"{o}rperelektronik, Technische
Universit\"{a}t Wien, Austria, and the Sub-micron center, Weizmann
Inst. of Science, Israel. We thank E.Gornik, C. Unterrainer and I.
Bar-Joseph of these institutions for their assistance. Last but
not least, we gratefully acknowledge P. Zoller and T. Calarco who
are responsible for the theoretical vision. This work was
supported by the Austrian Science Foundation (FWF), project
S065-05 and SFB F15-07, the Jubil\"aums Fonds der
\"Osterreichischen Nationalbank, project 6400, and by the European
Union, contract Nr. TMRX-CT96-0002. B.H. acknowledge financial
support form Svenska Institutet.


\begin{figure}
    \begin{center}\hspace{0mm}\mbox{\input epsf
\epsfxsize\columnwidth\epsfbox{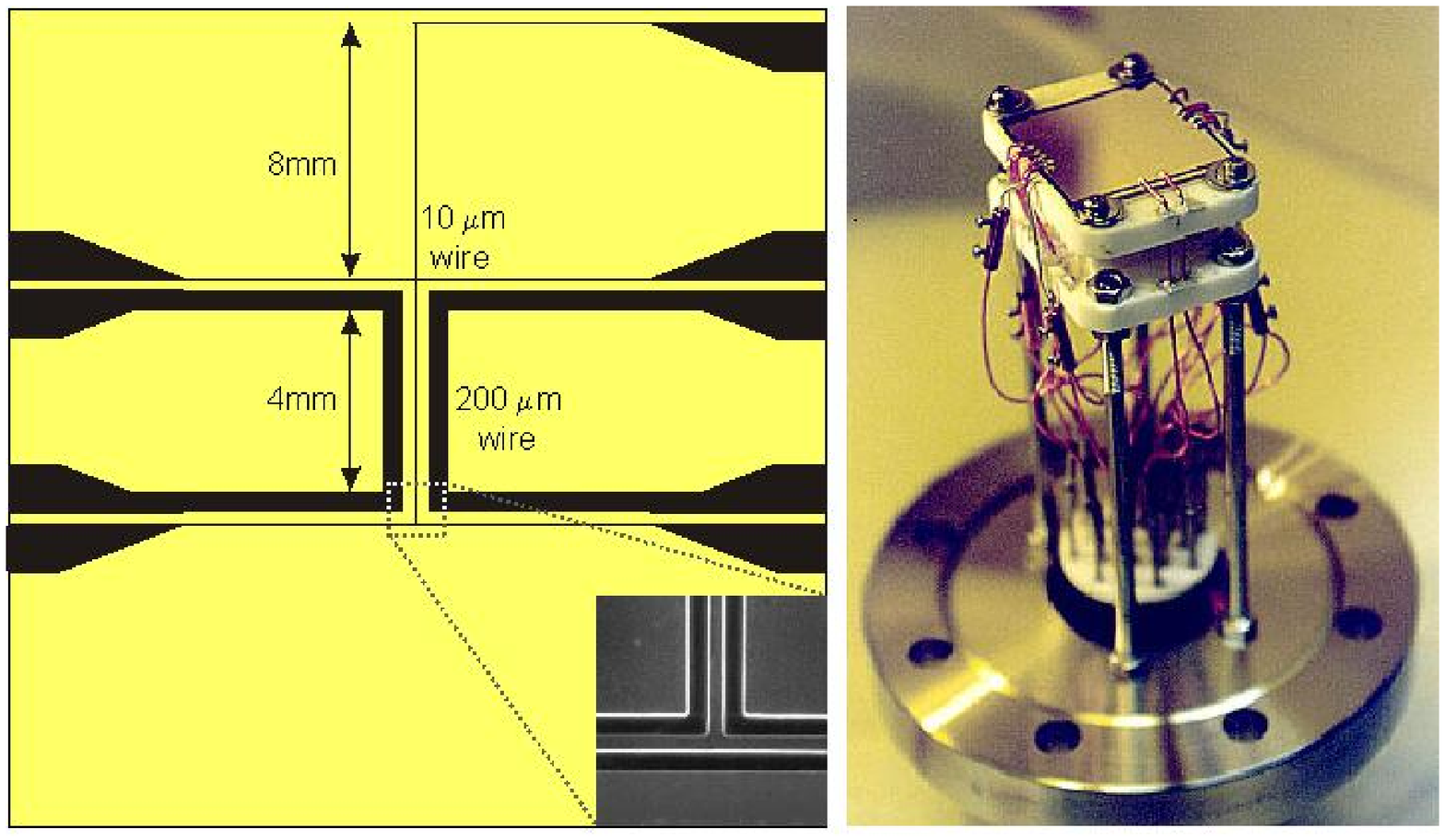}}\end{center}
    \vspace{1cm}
    \caption{(a) A schematic of the chip surface design.  For simplicity,
    only wires used in the experiment are shown. The wide wires are
    $200 \, \mu m$ wide while the thin wires are $10 \, \mu m$ wide.
    The insert shows an electron microscope image of the surface and
    its $10 \, \mu m$ wide etchings defining the wires. (b) The
    mounted chip before it is introduced into the vacuum chamber.}
    \label{fig:pict1}
\end{figure}

\newpage

\begin{figure}
    \begin{center}\hspace{0mm}\mbox{\input epsf
\epsfxsize0.8 \columnwidth\epsfbox{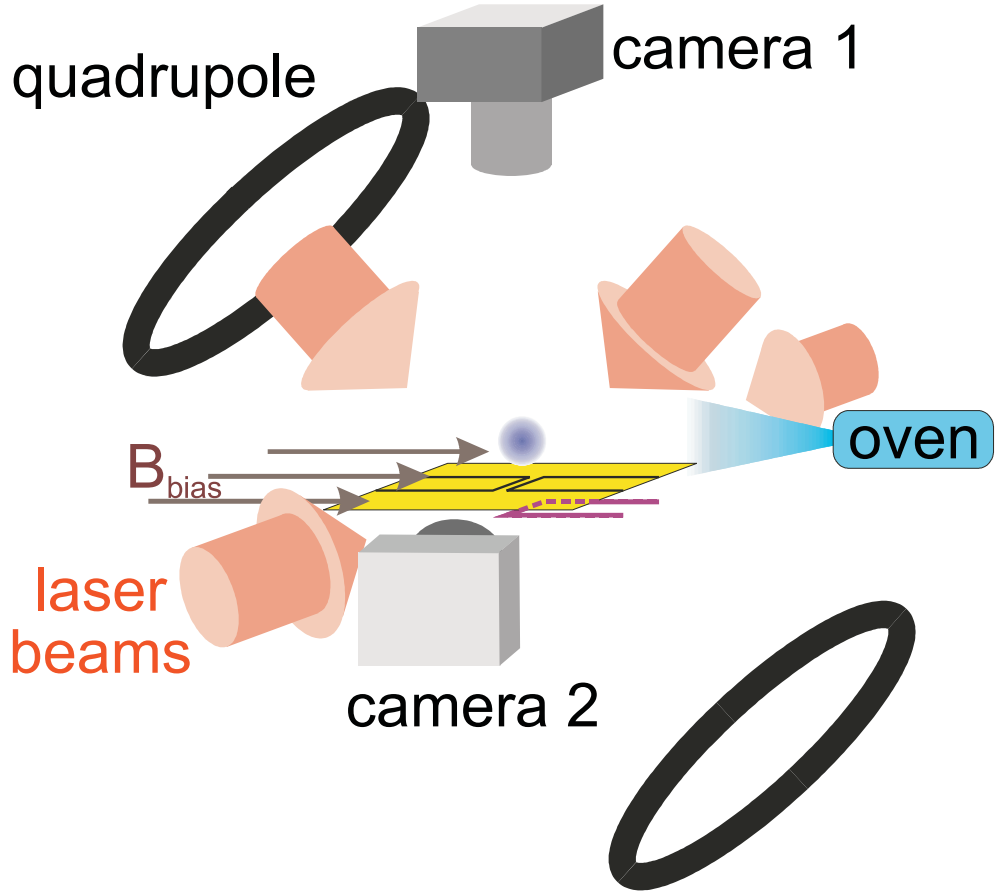}}\end{center}
    \vspace{1cm}
    \caption{Experimental setup: Four circularly polarized light
    beams enter the chamber; two are counter propagating parallel to
    the surface of the chip, while the two others, impinging on the
    surface of the chip at $45$ degrees, are reflected by the gold
    layer. The chip, the underlying U-wire
    trap, and the bias field, are oriented in such a manner as to
    provide a quadrupole field with the same orientations as the MOT
    coils. The oven, the effusive beam, and the two cameras observing
    the atomic cloud are also shown.}
    \label{fig:pict2}
\end{figure}

\newpage

\begin{figure}
    \begin{center}\hspace{0mm}\mbox{\input epsf
\epsfxsize0.5 \columnwidth\epsfbox{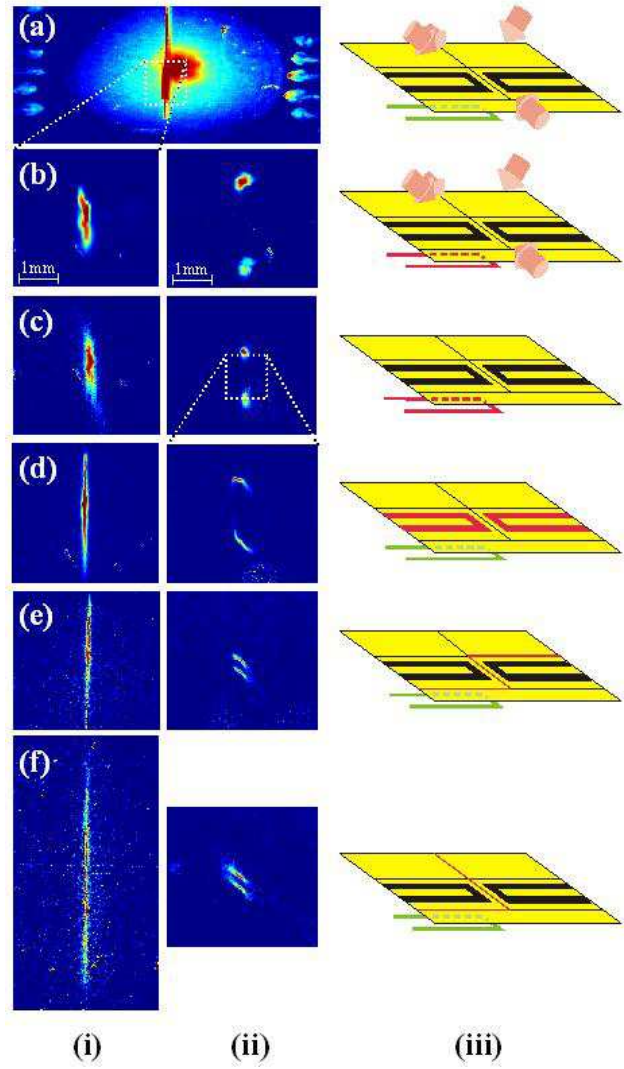}}\end{center}
    \caption{(color) Experiments with an Atom Chip: column (i) shows the view
    from the top (camera 1), column (ii) the front view (camera 2) and
    (iii) a schematic of the wire configuration.  Current carrying
    wires are highlighted in red.  The front view shows two images:
    the upper is the actual atom cloud and the lower is the reflection
    on the gold surface of the chip.  The distance between both images
    is an indication of the distance of the atoms from the chip
    surface.  Rows (a)-(f) show the various steps of the experiments.
    (a)-(d) show the step wise process of loading atoms onto the chip
    while (e) and (f) show atoms in a microscopic trap and propagating
    in a guide.  The pictures of the magnetically trapped atomic cloud
    are obtained by fluorescence imaging using a short laser pulse
    (typically $0.5 \, ms$)}
    \label{fig:pict3}
\end{figure}

\newpage

\begin{figure}
    \begin{center}\hspace{0mm}\mbox{\input epsf
\epsfxsize0.9 \columnwidth\epsfbox{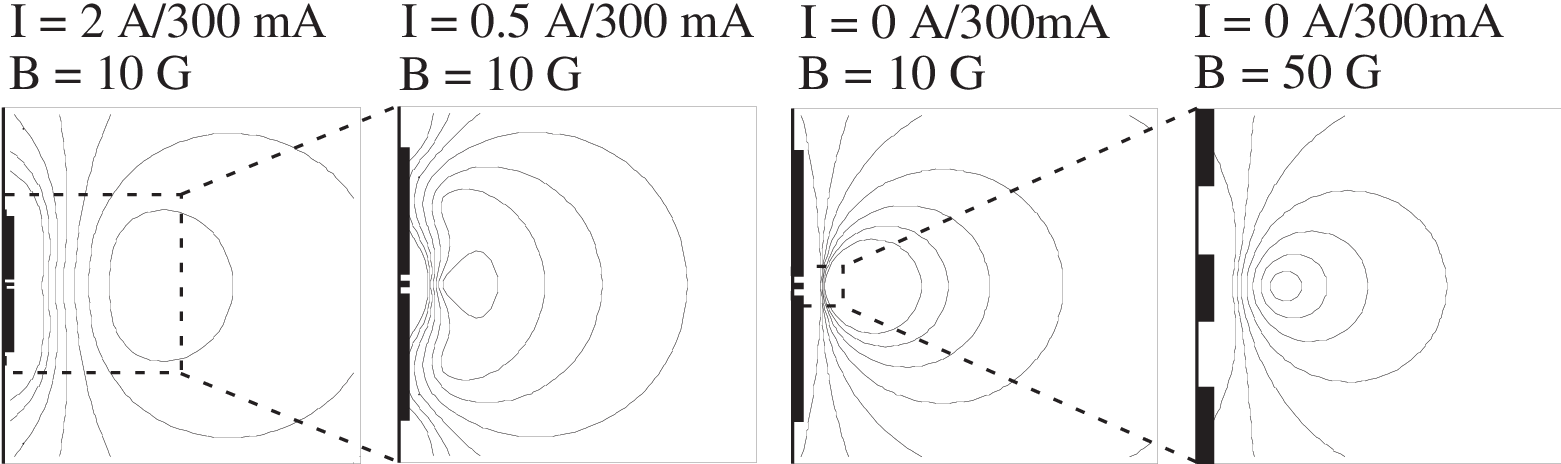}}\end{center}
    \vspace{2cm}
    \caption{Transfer from a large trap formed by two U-shaped wires
    to one thin wire:  The position of the surface mounted wires and
    equipotential lines for the trapping potential are shown.
    i) The first picture: the large 200 $\mu m$ U-traps carry a current
     of 2A and the small 10 $\mu m$ wire 300 mA.
    ii) The second picture shows an intermediate stage in the transfer
    to the 10 $\mu m$ trap. The current in the large U-traps is
    decreased to 0.5 A.
    iii) The large U-traps are now switched off and the transfer to
    the small 10 $\mu m$ trap is complete.
    iv) By increasing the bias field the trap can be compressed
    further.}
    \label{fig:compression}
\end{figure}


\begin{references}
\bibitem{LiMOT}
    The atoms are loaded into the MOT for $20s$ out of an effusive
    beam at a red laser detuning of $25MHz$ and a total laser power of
    about $150mW$. An electro-optic modulator produces sidebands of $812
    MHz (30\%)$ one of which is used as a repumper. To increase the
    loading rate we use an additional slower beam ($20mW$, $100MHz$ red
    detuned) directed through the MOT into the oven. The MOT is
    typically $1mm$ in diameter (FWHM) and has a temperature of
    T $\sim$  200 $\mu$K which corresponds to a velocity of about $0.5m/s$.
\end{references}
\end{document}